\documentclass[preprint]{revtex4}
\usepackage{epsfig,latexsym,amssymb}
\usepackage{latexsym}
\usepackage{amssymb}
\usepackage{amsfonts}
\usepackage{comment}
\usepackage{graphicx}
\usepackage{verbatim}

\newcommand{\be}{\begin{equation}}
\newcommand{\ee}{\end{equation}}
\newcommand{\bea}{\begin{eqnarray}}
\newcommand{\eea}{\end{eqnarray}}

\begin{document}

\title{Axion-like particle dark matter in the linear regime of structure formation}

\author{Qiaoli Yang$^{1,2,}$\footnote{qiaoli\_yang@hotmail.com} and Haoran Di$^{1,}$\footnote{haoran\_di@yahoo.com}}
\affiliation{$^{1}$School of Physics, Huazhong University of Science and Technology, Wuhan 430074, China\\$^{2}$Department of Physics, Jinan University, Guangzhou 510632, China}

\begin{abstract}
If axion-like particles (ALPs) constitute a major part of dark matter (DM), due to the bosonic nature and a relative small mass, they could behave differently from the point like dark matter particles on the formation of the cosmic structures. When studying the structure formation, it is often useful to consider DM as a special fluid with a given density and a given velocity. ALP fluid obeys a same continuity equation comparing to the point-like collisionless DM but has a different first-order velocity equation. In the linear regime of structure formation, the resulted observational differences are negligible for the QCD axions but can be interesting for very light ALPs.
\end{abstract}

\date\today
\maketitle

\section{Introduction}
The introduction of the Peccei-Quinn mechanism to explain the strong CP problem \cite{Peccei:1977hh} leads to a new pseudoscalar particle known as the axion \cite{Weinberg:1977ma, Wilczek:1977pj}. Generally speaking, a U(1) symmetry is introduced and spontaneously breaks down. A Goldstone boson appears and acquires a small mass from non-perturbative QCD effects thus it becomes a pseudo-Goldstone boson, the axion. Early axion models are quickly ruled out because their symmetry-breaking energy scales are not high enough to suppress their couplings to evade experimental detections. Subsequent models such as the KSVZ axions and the DFSZ axions \cite{Kim:1979if, Shifman:1979if,Zhitnitsky:1980tq,Dine:1981rt} have not been verified due to the very weak interactions between the axions and the standard model particles. Thus, these models are called the "invisible axions." Axion-like particles (ALPs) \cite{Svrcek:2006yi}\cite{Douglas:2006es} generically exist in the string theory, where antisymmetric tensor fields are compactified on closed cycles, and the respective Kaluza-Klein zero modes acquire non-zero potentials due to non-perturbative effects on the cycles \cite{witten:1986}\cite{Becker:1995kb}\cite{Kallosh:1995hi}. The potentials  give a small mass to the Kaluza-Klein modes and these Kaluza-Klein modes are called the ALPs since they have many properties similar to that of the axions. The parameter spaces of these ALPs, however, are much less constrained than that of the QCD axions \cite{Kaloper:2008qs}\cite{Arvanitaki:2009fg}.

Cosmological and astronomical observations indicate that the majority of matter in our universe is composed of unknown particle types. From observations such as the Cosmic Microwave Background (CMB) anisotropy and primordial nuclear synthesis, we know that these particles are non-baryonic, cold, weakly interacting and stable. The other properties of these dark matter particles remain unknown. Axions/ALPs produced by the mis-alignment mechanism and/or decay of cosmic topological defects can have the required properties to represent a substantial fraction of dark matter  \cite{Ipser:1983mw, Preskill:1982cy, Abbott:1982af, Dine:1982ah, Berezhiani:1990sy, Sikivie:2006ni}.

Currently, researchers are actively searching for cosmological or astrophysical original axions/ALPs, including ADMX \cite{Hoskins:2011a} (Univ. of Washington), CAST \cite{Arik:2014a} (Cern et al.). In contrast, PVLAS \cite{Kuster:2008zz} (Padova) and ALPS \cite{Ehret:2010mh} (DESY et al.) are aiming for the laboratory original axions/ALPs.

Although the major candidates for dark matter, such as axions/ALPs, WIMPs and sterile neutrinos et al., were thought to be indistinguishable by purely astronomical and cosmological observations, the axions/ALPs have two interesting characteristics: they are bosonic and are highly occupied in the phase space. In addition, when treated as a special fluid, the first order velocity equation of ALPs differs from that of the point-like dark matter particles, with additional terms that represent the quantum pressure and the self-interaction. The differences are negligible for the QCD axions in the linear regime, however, it is interesting to find whether these terms could have measurable consequences for extremely light ALPs with mass $m\sim 10^{-22}$eV. 

\section{Dark Matter Axions/ALPs}
The axion is a pseudo-Goldstone boson with a small mass due to the QCD instanton effect. The mass of an axion is determined by the potential term£º $V(a)=\Lambda^4[1-cos(a/f_a)]$, where $a$ is the axion field, $f_a$ is the symmetry-breaking scale and $\Lambda$ is the QCD scale factor of order $0.1{\rm GeV}$. From the axion potential, one can find the axion mass and self-interaction coupling by expanding near vacuum: $m=\Lambda^2/f_a$, $ {\cal L}_{int}=-{(\Lambda^4a^4)/ (4!f_a^4)}$. Let us denote the interaction term as $\lambda=-\Lambda^4/f^4_a$ for convenience. It is clear that the value of the symmetry-breaking scale $f_a$ is critical for axion phenomenology. The cosmological, astrophysical and experimental constraints on the value of $f_a$ are represented by $10^9{\rm GeV}<f_a<10^{12}{\rm GeV}$ \cite{Dicus:1978fp,Vysotsky:1978dc,Dicus:1979ch,Raffelt:1987yu,Dearborn:1985gp,Khlopov:1999tm,TheFermi-LAT:2016zue}.

The axions are not absolutely stable because, for example, they can decay into two photons, but their decay rate is much smaller than the Hubble rate. Thus, practically speaking, the axions are stable. For a discussion of the properties of dark matter axions, one may use the following Lagrangian density:
\be
\mathcal{L}={1\over2}(\partial a)^2-{1\over 2}m^2a^2-{\lambda\over 4!}a^4~~.
\ee

The cosmology axions have two population types. The thermal axions are produced by thermal equilibrium with the primordial soup, so that the number density is determined by the decoupling temperature $T_D$. Just before decoupling, we have \cite{Sikivie:2006ni}:
\be
n_a={1.202\over \pi^2}T_D^3({R_D\over R})~,
\ee
where $R_D$ and $R$ are the scale factors for decoupling and for today, respectively.
The decoupling temperature is determined by the condition $\Gamma>H$, where $\Gamma$ is the rate axions are created and annihilated. The thermal axion density is very small compared to the critical density.

The second population type is the cold axions. The cold axions can be created by two ways: vacuum realignment and the decay of axion-field topological defects, such as the axion strings and/or the axion domain walls if the inflation happens before the PQ symmetry breaking. If the inflation happens after the PQ symmetry breaking, the topological defects are homogenized so that only the vacuum realignment contributes. For both scenarios, the cold axions have an abundance of order \cite{Sikivie:2006ni}:
\be
\Omega_a\sim({f_a \over {10^{12} {\rm GeV}}})^{7/6}~,
\ee
which is a substantial fraction of the critical density if $f_a\sim 10^{12} {\rm GeV}$.

The ALPs are created due to compactifying antisymmetric tensor fields on closed cycles, and the respective Kaluza-Klein zero modes acquire a potential from non-perturbative effects on the cycles. There are many different ALPs due to the topology of the space-time manifold in the string theory and let us consider a scenario that one of them constitutes a substantial fraction of dark matter. The effective Lagrangian in the four dimensional space-time is:
\be
{\cal L}={f^2_{ALP}\over 2}(\partial a)^2-\Lambda_{ALP}^4U(a)+..~~,
\ee
where $f_{ALP}$ and $\Lambda_{ALP}$ are the decay constant and the potential energy scale for the ALPs, respectively. For small vibration of $a$ we can always expand $U$ around the minimum, so that the mass of the ALPs is of order $m\sim \Lambda_{ALP}^2/f_{ALP}$. The potential energy scale $\Lambda_{ALP}$ depends on the UV energy scale of the string theory and also is exponentially sensitive to string instantons \cite{Svrcek:2006yi}. Therefore, the mass range of the ALPs is distributed on a large scale.

The ALPs DM can be created in the early universe in a similar way as the QCD axions. The equation of motion of a homogenized ALPs field in a flat Friedmann-Robertson-Walker (FRW) universe is:
\be
\partial^2_ta+3H\dot a+m^2a=0~,
\label{axioninfrw}\ee
where $H$ is the Hubble rate. After the Hubble rate decreases to comparing to the mass of the ALPs at time $t_1$, $t_1\sim 1/m$, the ALPs field starts to oscillate and the initial oscillation amplitude is determined by the vacuum realignment angle. The newly created ALPs are very cold so the energy density decreases as $\rho\propto R^{-3}$. The energy density of ALPs today is therefore $\rho\sim {(1/2)}m_a^2f_a^2[R(t_1)/R(t)]^3$ and the ratio of the ALPs' energy density to the critical energy density $\rho_c$ today is: $\rho/\rho_c\sim 3\pi Gf_a^2\sqrt{mt_e}\lesssim 1$, where $t_e$ is the matter-radiation equality time. In addition, the ALPs DM should be stable, which requires that the decay rate: $\Gamma\approx \alpha^2 m^3/(4\pi^3f_a^2)\lesssim H_0$, where $H_0$ is the Hubble rate today and $\alpha$ is the fine structure constant. The above two considerations give an upper limit of the mass of the DM ALPs: $3\pi G m^{7/2}{t_e}^{1/2}(\alpha^2/4\pi^3)\lesssim H_0$, which implies $m\lesssim 30$keV. Finally, as the Compton wavelength of the DM ALPs should be smaller than 1kpc to allow the structure formation on the kpc scale, the DM ALPs have a lower limit on the mass $m\gtrsim 10^{-24}$eV.

\section{Axion/ALP DM Comparing to the point-like collisionless DM}
After the cold axions/ALPs are created by the vacuum realignment mechanism, these dark matter particles have a very high number density and a very small velocity dispersion. The high number density combined with a small velocity dispersion results in a very high phase space density. The phase space density $\cal N$ of the QCD axion DM is \cite{Sikivie:2009qn}:
\be
{\cal N}\sim 10^{61}({f_a\over 10^{12}{\rm GeV}})~.
\ee
For the ALPs lighter than the QCD axions, the phase space density is even higher.

Because the quantum states is highly occupied by particles, one may use approximation where operators become classical fields. Let us consider the axions/ALPs' field in an unperturbed flat FRW universe, the Lagrangian density is:
\be
{\cal L}={1\over2}\dot a^2R^3-{R\over 2}(\nabla a)^2 -{1\over 2}m^2a^2R^3-{\lambda\over 4!}a^4R^3+..~.
\ee
The equation of motion is thus:
\be
\partial^2_ta-{1\over R^2}\nabla^2 a+3H\partial_ta+m^2a+{\lambda\over6}a^3=0~~,
\ee
where $H=\dot R/ R$ is the Hubble parameter.

The cold axions/ALPs are non-relativistic, so the slow varying term of the field is most important while the rapid varying terms of order axion/ALPs mass, such as $e^{\pm imt}$, can be factored out. Expressing that $a=(\psi e^{-imt}+\psi^*e^{imt})/\sqrt {2m}$, where $\psi$ is a complex scalar, the equation of motion for the slow varying term $\psi$ is:
\be
-i\dot \psi-{1\over 2m R^2}\nabla^2\psi-i{3\over 2}H\psi+{\lambda\over 8m^2}|\psi|^2\psi=0~.
\label{waveeq}
\ee
To comparing the equation of motion with the point-like collisionless DM, it is useful to re-express Eq.~(\ref{waveeq}) in terms of density and effective velocity. In comoving coordinates, one finds the generalized continuity equation:
\be
{\partial n_a\over \partial t}+{1\over R}{\partial(n_a v^i)\over \partial x^i}+3Hn_a=0~,
\label{density}
\ee
where $n_a\equiv \psi\psi^*$ is the particle density and $v^i\equiv -i(\psi^*\partial_i \psi-\psi\partial_i\psi^*)/(2mRn_a)$ is the effective velocity. The continuity equation is the same as that of the point-like DM. However, the first-order velocity equation is different:
\be
{\partial v^i\over \partial t}+Hv^i+{\lambda \over 8m^3R}\partial_i n_a-{1\over 2m^2R^3}\partial_i({{\nabla}^2\sqrt n_a\over \sqrt n_a})=0~,
\label{velocity}
\ee
comparing with that of the point-like DM
\be
{\partial v^i\over \partial t}+Hv^i=0~~.
\ee
There are two additional terms: $\lambda\partial_i n_a/( 8m^3R)$ and $-{(1/ 2m^2R^3)}\partial_i[{({\nabla}^2\sqrt n_a)/ \sqrt n_a}]$. The first one comes from the self-interaction and the second one comes from the quantum pressure.

For the QCD axions, the self-interaction term and the quantum pressure term are too small to be considered for observable scales. This is obvious if one considers the Jeans length due to the self-interaction and the quantum pressure:
\be
\partial^2_t\delta+2H\partial_t\delta+({k^4\over4m^2R^4}-4\pi G\bar \rho-{4\pi\sigma^{1/2}\bar\rho\over m^3}{k^2\over R^2})\delta=0~,
\ee
where $\delta=\delta \rho/\bar\rho$ is the density contrast, $\bar\rho$ is the averaged density of DM and $\sigma$ is the cross section of the self-interaction. For QCD axions in the linear regime of density fluctuation, only very small-scale structures, $(k/R)^{-1}\sim (\sqrt\sigma/Gm^3)^{1/2}$ and $(k/R)^{-1}\sim (16\pi m^2G\rho)^{-1/4}$, differ from the point-like collisionless particle DM (see FIG.1). However, for the ALPs, the additional terms in Eq.(\ref{velocity}) can be important if the mass is sufficiently small.

\begin{figure}
\begin{center}
\includegraphics[width=0.8\textwidth]{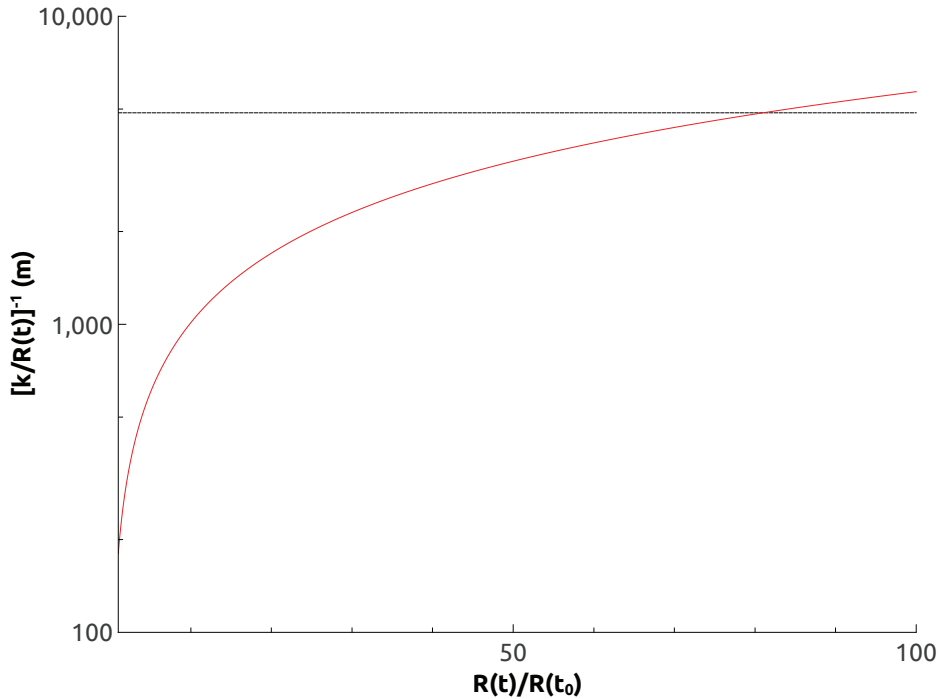}
\caption{The quantum pressure versus the gravity (the quantum pressure dominates over the gravity below the red-line), and the self-interaction versus the gravity (the self-interaction dominates over the gravity below the black-dashed line) of the QCD axions. Thus in the linear regime of structure formation, the QCD axion DM behaves differently from the point-like collisionless particle DM only for very small scales $\sim 10$km.}
\end{center}
\end{figure}

In the non-linear regime of the evolution of axions/ALPs, the DM axions/ALPs thermalize due to the gravity interaction \cite{Sikivie:2009qn}, thus form a Bose-Einstein Condensation (BEC). If the wave function of the axions/ALPs can be written as:
\be
\psi=\sqrt{{n_a(\vec x,t)}}e^{i\phi(\vec x,t)},
\ee
and the effective velocity is defined as $v\equiv \partial_i\phi/(Rm)$, the Landau's postulates of a superfluid is then a natural consequence of the axions/ALPs BEC. In the non-linear regime of structure formation, superfluid behave very differently from point-like DM. For example, as $\nabla\times \vec v=0$, the angular momentum of the dark matter halos needs to be accommodated by appearance of quantized vortices. A detailed discussion of the formation of vortices in DM haloes is given in \cite{RindlerDaller:2011kx}.

\section{Summary}
We consider a scenario that the majority of dark matter is composed by axion like particles. The cosmological and astrophysical considerations give rise to a mass constraint for the ALPs $10^{-24}{\rm eV}\lesssim m \lesssim30{\rm keV}$. In addition, the first order velocity equation of the ALP fluid differs from that of the point-like collisionless DM particles with two additional terms that represent the quantum pressure and the self-interaction of ALPs respectively. The resulted differences in the linear regime of structure formation are negligible for the QCD axions. However, it remains interesting to find whether these terms have measurable consequences for extremely light ALPs with mass $m\lesssim  10^{-22}$eV.

{\itshape Acknowledgments:} We would like to thank Weitian Deng, Bo Feng and Jianwei Cui for their helpful discussions. This work is partially supported by the Natural Science Foundation of China under Grant Number 11305066.

\end{document}